\documentclass[aip,reprint,twocolumn]{revtex4-1}
\usepackage{graphicx,amssymb, amsmath,framed,amsthm}
\usepackage[caption=false]{subfig}

\usepackage{mathtools}
\usepackage{amsfonts}
\usepackage{tikz}
\usetikzlibrary{patterns}

\usepackage{subfig}
\DeclareMathAlphabet\mathbfcal{OMS}{cmsy}{b}{n}

\makeatletter
\@ifpackageloaded{stix} 
{
	
}{                      
	\usepackage{mathrsfs}
	\usepackage{txfonts}              
}
\makeatother

\usepackage[unicode=true,pdfusetitle,bookmarks=true,bookmarksnumbered=false,bookmarksopen=false,breaklinks=false,pdfborder={0 0 0},backref=false,colorlinks=true]{hyperref}
\hypersetup{citecolor=blue,filecolor=blue,linkcolor=blue,urlcolor=blue}
\usepackage{color,soul}
\usepackage{siunitx} 
\usepackage{graphicx}
\usepackage{dcolumn}
\usepackage{cancel}
\usepackage{diagbox} 
\usepackage{multirow}

\newcommand{\p}{\partial}
\newcommand{\f}[2]{\frac{#1}{#2}}

\newcommand{\lf}{\left}
\newcommand{\ri}{\right}

\newcommand{\delf}[2]{\frac{\delta #1\hfill}{\delta #2\hfill}}

\makeatletter
\newcommand{\vast}{\bBigg@{4}}
\newcommand{\Vast}{\bBigg@{5}}
\makeatother

\newcommand{\calH}{\mathcal{H}}

\newcommand{\calP}{\mathcal{P}}

\DeclareMathAlphabet{\mathpzc}{OT1}{pzc}{m}{it}

\newcommand{\teb}{\textcolor{blue}}

\begin{document}
	\title{Direction of cascades in a magnetofluid model with  electron skin depth and ion sound Larmor radius scales}
	\author{George Miloshevich}
	\email{milosh@utexas.edu}
	\affiliation{Department of Physics and Institute for Fusion Studies, The University of Texas at Austin, Austin, TX 78712, USA}
	\author{Philip J. Morrison}
	\email{morrison@physics.utexas.edu}
	\affiliation{Department of Physics and Institute for Fusion Studies, The University of Texas at Austin, Austin, TX 78712, USA}
	\author{Emanuele Tassi}
	\email{tassi@cpt.univ-mrs.fr}
	\affiliation{Aix Marseille Univ, Univ Toulon, CNRS, CPT, F-13288 Marseille cedex 09, France}
\affiliation{ Universit\'e C\^ote d'Azur, CNRS, Observatoire de la C\^ote d'Azur,
Laboratoire J.L. Lagrange, Boulevard de l'Observatoire, CS  34229, 06304 Nice Cedex 4, France}
	
\date{\today}

\begin{abstract}
The direction of cascades in a two-dimensional model that takes electron inertia and ion sound Larmor radius into account is studied, resulting in  analytical expressions for the absolute equilibrium states of the energy and helicities. These states suggest that typically both the energy and magnetic helicity at scales shorter than electron skin depth have direct cascade, while at large scales the helicity has an inverse cascade as established earlier for  reduced magnetohydrodynamics (MHD). The calculations imply  that the introduction of gyro-effects  allows for the existence of negative temperature {(conjugate to energy)} states and the condensation of energy to the large scales. Comparisons between two- and three-dimensional extended MHD models (MHD with two-fluid effects) show  qualitative agreement between the two.
\end{abstract}
\maketitle

\section{Introduction}
\label{intro}

{
In various astrophysical and laboratory settings, plasma is known to be in a turbulent state. Progress in the understanding of turbulence is thus crucial for explaining several phenomena occurring in astrophysical and laboratory plasmas. On sufficiently large scales, magnetohydrodynamics (MHD)  is a valid model for describing plasma turbulence and is indeed the basis for the theoretical descriptions of several plasma phenomena. Among these, for instance, the magnetic dynamo action (see, e.g.\  Ref.~\onlinecite{Bis00}), which has been established as a mechanism for conversion of kinetic energy into magnetic energy. Such conversion is relevant for the Earth's magnetosphere as well as the solar wind. The dynamo action has also been linked to the inverse cascade in MHD turbulence.\cite{frisch75, fyfe76, fyfe77, montgomery79,brandenburg01} Theoretical predictions for MHD turbulence have been confirmed in numerical simulations\cite{seyler75,fyfe77} and similar works have been successfully undertaken in three-dimensional (3D) Hall MHD.\cite{servidio08} The magnetic relaxation process characterizing magnetically confined plasmas in Reversed Field Pinches \cite{Ort93} is another example of phenomenon whose understanding is based on the MHD description of a turbulent plasma. Further applications of MHD turbulence can be found, for instance, in Ref.~\onlinecite{biskamp03}.}

{While MHD has been a cornerstone for the description of  large scale plasma phenomena, it fails at short scales, such as the electron skin depth  $\hat{d}_e= c/\omega_{pe}$, where  $c$ is the speed of light and $\omega_{pe}$ is the electron plasma frequency. The model of extended MHD (XMHD) generalizes MHD (as well as Hall MHD) by including  terms that are relevant at scales of the order of $\hat{d}_e$. The investigation of turbulence at such scales is of relevance for instance for the  recently launched Magnetospheric Multiscale Mission,\cite{burch16} which is known to be capable of probing such scales (observational results in these regimes have been recently published in Ref.~\onlinecite{narita14}). The probing of such scales may also become feasible in the laboratory, with facilities such as WiPAL.\cite{forest15}  The direction of turbulent cascades in three-dimensional (3D) XMHD was investigated in Ref.~\onlinecite{miloshevich17}.}

{Besides Hall MHD and XMHD a number of reduced fluid models  exist that account for additional two-fluid plasma effects, models that  are amenable to simpler analytical and numerical treatments. Such reduced models (see, e.g.,  Ref.~\onlinecite{Tas17}) typically rely on the assumption of a magnetic field with a strong constant component along one direction (strong guide field assumption) and are valid at frequencies much lower than the ion cyclotron frequency based on such a guide field. This situation is relevant for some laboratory plasmas as well as for a number of astrophysical situations (see, e.g.,  Ref.~\onlinecite{schekochihin09}). These  models are also characterized by the property of possessing only quadratic nonlinearities and by a spatial anisotropy induced by the presence of the guide field.}

{The purpose of this paper is to investigate the direction of turbulent cascades in one such reduced fluid model\cite{cafaro98} that accounts for  the electron skin depth scale and an additional scale consisting of the ion sound Larmor radius  $\hat{\rho}_s=\sqrt{T_e/m_i}/\omega_{ci}$, with $T_e$ the equilibrium electron temperature, $m_i$ the ion mass and $\omega_{ci}$ the ion cyclotron frequency based on the guide field.   This additional  scale, which accounts for finite electron temperature, proved, for instance, to be crucial for the nonlinear evolution of the current density and plasma vorticity during  a magnetic reconnection  process.\cite{Gra01,Del06}} 

{In our analysis we consider the two-dimensional (2D) case, assuming translational invariance along the direction of the guide field.  This assumption could be justified by the presence of  a strong guide field. We remark that, in its original and more general formulation,\cite{schep94}  the model assumes   only weak variations along the guide field and, in particular, nonlinear terms only involve derivatives along directions perpendicular to the direction of the guide field. Moreover, the appearance of coherent structures in  two-dimensional turbulence and the possible occurrence of reconnection events induced by electron inertia, as suggested by recent observations of the fast solar wind,\cite{Per17}  make the 2D version of the model of interest in its own right. Some comparison with cascade properties of 3D XMHD will nevertheless be made. In 2D the model is known to possess two infinite families of  integral invariants \cite{schep94} (Casimir invariants) associated with the noncanonical Hamiltonian structure of the model. A qualitative change in the form of these families of  invariants occurs when the normalized ion sound Larmor radius $\rho_s=\hat{\rho}_s/L$ (with $L$ a characteristic length of the system) is set equal to zero.}


  

{In order to predict the direction of turbulent cascades of invariants of the model, we resort to the well-known technique of absolute equilibrium states (AES).}
This technique  (see Sec.~\ref{AES}) has been used  in various past works: it was  applied to hydrodynamical turbulence in  Ref.~\onlinecite{kraichnan73}, MHD in Refs.~\onlinecite{frisch75,fyfe76,fyfe77}, Hall MHD in Refs.~\onlinecite{servidio08,miloshevich17}, two-fluid theory   in Ref.~\onlinecite{zhu14}, 3D XMHD in Ref.~\onlinecite{miloshevich17}, gyrokinetics in Ref.~\onlinecite{zhu10},  {and drift wave turbulence in Refs.~\onlinecite{gang89,koniges91}}.  AES are derived from the Gibbs ensemble probability density and represent  states towards which  actual turbulent  tends to relax;  thereby,  they are of value for  predicting the direction and structure of the exchange of various invariants among the modes.\cite{kraichnan80}  It is important to  mention that these modes are not  eigenstates of the  various models considered, but Fourier amplitudes that  allow analyses of how components of the invariants flow through different  scales. 



One of the earliest suggestions for ascertaining the inverse cascade based on AES  in MHD turbulence\cite{biskamp03} can be found in Ref.~\onlinecite{frisch75},  followed by the two-dimensional studies of Ref.~\onlinecite{fyfe76},  inspired by works of Kraichnan in hydrodynamics.\cite{kraichnan73,kraichnan80} Numerical simulations \cite{fyfe77}  support the predicted relaxed spectra.  Although later it was found that deviation from Gaussian statistics occurs as well as breaking of ergodicity in MHD.\cite{shebalin89}  Good agreement between AES in Hall MHD and numerics was found {in Ref.}~\onlinecite{servidio08}.  Later mostly analytical calculations for AES were performed in two-fluid theory \cite{zhu14} and gyrokinetics,\cite{zhu10} where the  former alludes to the possibility that the {{``poles''} of AES can appear in the high-$k$ regime and a pole implies condensation of a spectral quantity to that wavenumber (see Sec.~VI)}. More detailed analyses were performed in Ref.~\onlinecite{miloshevich17},  predicting the phenomenon of cascade reversal of the magnetic helicity in 3D extended MHD at the electron skin depth scale. An in-depth overview can be found in Refs.~\onlinecite{shebalin13,miloshevich17}. 


\textcolor{black}{One of the main objectives of the present analysis is  to see if the cascade reversal of magnetic helicity at the electron skin depth predicted in Ref.~\onlinecite{miloshevich17}  has a counterpart in the 2D reduced model considered in this paper. (We anticipate that, when neglecting toroidal velocity and magnetic field components, the 2D incompressible limit of XMHD,\cite{grasso16} which we will refer to as to '2D planar incompressible XMHD', formally reduces to the 2D reduced model studied here in the limit $\rho_s=0$). As is well known, the directions of cascades change when going from 3D to 2D in hydrodynamics and in MHD, although in the latter case regimes exist where AES predict the same direction for energy cascade in 2D and 3D (see, e.g., Ref.~\onlinecite{Bis93}). }

{The identification of cascade reversal  is a subject that has attracted  considerable interest. However, mostly cascade reversals (usually referred to as cascade transitions in the literature) have only been identified in  highly idealized systems. For instance, there are many examples of  cascade reversal when the interactions of the real physical system have been artificially modified.}

 For example, in Ref.~\onlinecite{sahoo17}  it is demonstrated that 3D hydrodynamics (HD) displays a change in the direction of the energy cascade when varying the value of a free parameter that controls the relative weights of the triadic interactions between different helical Fourier modes. Another useful study was  performed in a model of thin layer turbulence, \cite{benavides17} where 2D motions were coupled to a single Fourier mode along the vertical direction. As the height of the layer is varied the authors find critical transitions from forward to backward cascade of energy. 
	
{The literature on  cascade reversal in real physical systems, ones without artificial modification,  is scarcer. 
Some examples include rotating three dimensional stirred HD system,\cite{smith99,Deusebio14} where the transition to inverse cascade of energy occurs below certain values of Rossby number. In addition,  in Ref.~\onlinecite{Pouquet13}   3D direct numerical simulations of rotating Boussinesq turbulence also demonstrate such transitions. There is also theoretical and experimental evidence for the inverse energy cascade in the second sound acoustic turbulence in superfluid Helium.\cite{ganshin08} In 3D MHD,  various simulations have been performed \cite{alexakis16,sujovolsky16} that demonstrate cascade reversal when the system is forced only mechanically. Since the stirring lacks a magnetic component with stronger guide fields the flow becomes two-dimensional,  leading to the inverse cascade of energy like in 2D HD. The transition appears to have some interesting features \cite{Seshasayanan14} as the magnetic forcing is turned on, viz.\  there exists a critical value for which the energy flux towards the large scales vanishes. In the MHD examples above, the parameter that is varied corresponds to the form of the amplitude
of a magnetic forcing added to the MHD system. {In our paper, on the other hand, the possible occurrence of cascade reversal will be investigated adopting $d_e$ and $\rho_s$ as control parameters.} }

 {The paper is organized as follows. We review the reduced fluid  model and its Hamiltonian structure in Sec.~\ref{model}, while a  discussion of its spectral decomposition properties follows in Sec.~\ref{spec}.  In Sec.~\ref{AES} we present  our calculations of AES, whereas in Sec. \ref{sec:regimes} we discuss the different regimes that characterize the AES depending on the values of the parameters. Finally, in Sec.~\ref{comparisons} we discuss comparisons with other related models and summarize.   }

\section{The model and its invariants}
\label{model}

{As stated in Sec.~\ref{intro}, we consider the model of Ref.~\onlinecite{cafaro98}, which was used earlier in Hamiltonian reconnection {studies}.
{\cite{grasso99,Gra01}} This model is applicable to low-$\beta$ plasmas, with $\beta$ indicating the ratio of the kinetic and magnetic pressures, and it can be seen as an extension of the previously investigated reduced MHD model of Ref.~\onlinecite{fyfe76}, accounting also for the effects of electron inertia and finite, constant electron temperature. As such, it describes plasmas with a strong magnetic guide field and it can be used to  locally  model  phenomena such as collisionless reconnection and turbulence, in situations where a detailed description of the temperature and heat flux evolution is {not required}.} Because the processes occur on time scales shorter than dissipation time scales, a collisionless Hamiltonian treatment is appropriate. However,  in a realistic turbulence scenario   dissipation cannot be ignored, even if the resistivity and viscosity appear negligible. The model can be obtained from a more general three-field model \cite{schep94} in the cold ion limit {and assuming an ion response with ion density fluctuations proportional to vorticity fluctuations.  Alternatively, the model can  be obtained from a  two-moment closure of drift-kinetic equations.\cite{deB01,Zoc11,Tas15}}

{The model equations, in dimensionless form, are given by 
\begin{equation} \label{2dxmhdrealequations}
\begin{split}
\frac{\partial \psi^\star}{\partial t} &=\{\psi^\star,\mathcal{H}\}=  \lbrack  \psi^\star, \phi \rbrack +\rho_s^2\lbrack \omega,\psi\rbrack,\\
\frac{\partial \omega}{\partial t} &=\{\omega,\mathcal{H}\}=  \lbrack  \omega, \phi \rbrack + \lbrack  \psi^\star, \nabla^2\psi\rbrack ,
\end{split}
\end{equation}
where $\omega = \nabla^2 \phi$ indicates the vorticity associated with a stream function $\phi$ (normalized electrostatic potential), whereas $\psi^\star = \psi - d_e^2 \nabla^2\psi$,
with $\psi$ the poloidal magnetic flux function of a magnetic field $\mathbf{B}=\nabla \psi \times \hat{z}+\hat{z}$.  The parameter   $d_e$ denotes the constant electron skin depth and the  second constant parameter $\rho_s$ corresponds  to the ion sound Larmor radius. The bracket $[\,,\,]$ is defined  as usual by  $\lbrack f, g\rbrack := \nabla f\times \nabla g\cdot \hat{z}$ for two functions $f$ and $g$ and the noncanonical Poisson bracket $\{\,,\,\}$ is defined below.}

{Using a caret to denote dimensional quantities, we have adopted the normalizations, $d_e=\hat{d}_e/L$,  $\rho_s=\hat{\rho}_s /L$, $t=\hat{t}/\tau_A, \phi=c \hat{\phi}/(B_0 v_A L)$, and $\psi=\hat{\psi}/(B_0 L)$, where as noted above $L$ is a characteristic length  and  $\tau_A=L/v_A$ with  $v_A$ being the Alfv\'en speed based on the amplitude $B_0$ of the guide field. The latter is assumed directed along the $\hat{z}$ axis of a Cartesian coordinate system $(x,y,z)$. Due to the 2D assumption, the $z$ coordinate is taken as ignorable. Note that, when two-fluid effects are suppressed (i.e. $d_e=\rho_s=0$), the model reduces to the 2D reduced MHD model of Ref.~\onlinecite{Str76}.}
  
The first equalities of Eqs.~(\ref{2dxmhdrealequations}) indicate that the system possesses a Hamiltonian formulation characterized by a Hamiltonian functional
\begin{equation}
\label{ham}
\mathcal{H} := \frac{1}{2} \int d^2x\,(-\phi \,\omega - \psi^\star \nabla^2 \psi +  \rho_s^2\omega^2),
\end{equation}
and  a noncanonical Poisson bracket (see Ref.~\onlinecite{morrison98} for review) 
\begin{eqnarray}
\label{RXMHDbracket1}
\{P, Q\} &=& \int d^2x\lf\{ \omega\lf( \lf[\delf{P}{\omega},\delf{Q}{\omega} \ri] +  \teb{d_e^2 \rho_s^2}\lf[\delf{P}{\psi^\star},\delf{Q}{\psi^\star} \ri] \ri)
\nonumber\ri.\\
&+&\lf. \psi^\star \lf(\lf[\delf{P}{\psi^\star},\delf{Q}{\omega} \ri]+ \lf[\delf{P}{\omega},\delf{Q}{\psi^\star} \ri]\ri) \ri\}\,.
\end{eqnarray}


{We remark that when electron temperature effects are neglected, i.e., when $\rho_s=0$, Eqs.~(\ref{2dxmhdrealequations}) reduce to the 2D inertial MHD (IMHD) system of Ref.~\onlinecite{lingam15dec} or, equivalently, as stated above,  to 2D planar incompressible XMHD}.

 The complexity of the Poisson Bracket of \eqref{RXMHDbracket1} can be reduced by the  coordinate transformation $\psi_\pm := \psi^\star \pm d_e\rho_s \omega$  to normal coordinates, in which the Poisson bracket has the following form:
\begin{equation}\label{PBSPKmodel}
	\{P, Q\}=  \!2 d_e \rho_s\!\!\int \!\!d^2x\lf(\psi_-\lf[\delf{P}{\psi_-},\delf{Q}{\psi_-} \ri] - \psi_+ \lf[\delf{P}{\psi_+},\delf{Q}{\psi_+} \ri]\ri).
\end{equation}
With the  bracket in the form of  \eqref{PBSPKmodel},   it is easily seen that the systems possesses two {infinite} families of Casimir invariants:
\begin{equation}\label{casimirs}
{C_{\pm}} = \int d^2 x \, \mathcal{F}_\pm (\psi_\pm)\,, 
\end{equation}
for arbitrary functions $\mathcal{F}_\pm$. {Casimir invariants are} functionals $C$ that satisfy $	\{C, Q\}=0$ for all functionals  $Q$. They are thus preserved for  dynamics generated by any Hamiltonian.

\section{Spectral Analysis}
\label{spec}

Equilibrium states of XMHD have been studied (see, e.g., Ref.~\onlinecite{kaltsas17}) leading to a generalization of  the Grad-Shafranov equation.  In contrast,  here we are interested in statistical equilibrium in  Fourier space, and the analysis of the associated direction of cascades.  In order to apply equilibrium statistical mechanics to the Fourier series of this system one has to prove a Liouville theorem  to ensure that a measure is conserved.\cite{landau1980statistical} Various systems have been shown to possess such,  including hydrodynamics, MHD,  and extended MHD in 3D.\cite{burgers29a,lee52,kraichnan80,miloshevich17} 

Using  a  standard Fourier representation $\psi(\mathbf{x}) = \sum_{\mathbf{k}} \psi_\mathbf{k} \,e^{i\, \mathbf{k}\cdot\mathbf{x}}$, so that $\psi_\mathbf{k}^\star = (1+k^2 d_e^2)\psi_\mathbf{k}$, Eqs.~\eqref{2dxmhdrealequations} become  
\begin{equation}\label{fourierpsieq}
	\dot \psi_\mathbf{k}^\star =  \hat{z}\cdot\sum_{\mathbf{k}^\prime,\mathbf{k}^{\prime\prime}} \delta_{\mathbf{k},\,\mathbf{k}^\prime+\mathbf{k}^{\prime\prime}}\,{\mathbf{k}^{\prime\prime}\times\mathbf{k}^{\prime}}\, \lf(\f{\omega_{\mathbf{k}^\prime}\, \psi_{\mathbf{k}^{\prime\prime}}^\star}{k^{\prime\, 2}} + \rho_s^2\, \omega_{k^\prime}\psi_{k^{\prime\prime}}\ri)
\end{equation}
and
\begin{equation}\label{fourieromegaeq}
\dot \omega_\mathbf{k} = \hat{z}\cdot\sum_{\mathbf{k}^\prime,\mathbf{k}^{\prime\prime}} \delta_{\mathbf{k},\,\mathbf{k}^\prime+\mathbf{k}^{\prime\prime}}\,\mathbf{k}^{\prime\prime}\times\mathbf{k}^{\prime}\, \lf(\f{\omega_{\mathbf{k}^\prime}\, \omega_{\mathbf{k}^{\prime\prime}}}{k^{\prime\,2}} + k^{\prime\, 2} \psi_{\mathbf{k}^\prime}\, \psi_{\mathbf{k}^{\prime\prime}}^\star\ri)\,.
\end{equation}
These equations can be generated by the Hamiltonian of \eqref{ham} and Poisson bracket of \eqref{RXMHDbracket1}  written in terms of Fourier series.   Consequently, they preserve the energy and all Casimir invariants  written in terms of their Fourier series. 

Of particular interest to us are the quadratic invariants preserved by \eqref{fourierpsieq} and \eqref{fourieromegaeq}, the so-called rugged invariants.  These are the Hamiltonian and the quadratic Casimirs.   The main reason for this is that such invariants survive  wave-number truncations,  $k_{min} < k < k_{max}$, which is common for  spectral Galerkin codes.  
Another  motivation  for using these invariants is  the ease of handling Gaussian statistics.

Of course, in general there may be other criteria, possibly motivated by experimental results, to ignore or select certain invariants in an  analysis of our type, based on the effects of viscosity/resistivity or other aspects ignored in ideal models. For instance in order to determine the relevant invariants,  the authors of {Ref.}~\onlinecite{jung06} have resorted to experiments. In our case, this possibility is excluded by the difficulty of obtaining experimental measures on the invariants for our system.  Therefore
we  stick with the quadratic invariants 
and  introduce linear combinations of the Casimirs  of  \eqref{casimirs}, viz.\ the following:
\begin{eqnarray}
F&:=& \frac{1}{2} \int {d^2} x \,\lf[\lf(\psi^\star\ri)^2 + d_e^2\rho_s^2 \, \omega ^2 \ri]
\label{FQcon}\\
G&:=& \int {d^2} x  \, \omega\, \psi^\star\,.
\label{GQcon}
\end{eqnarray}
The Hamiltonian \eqref{ham} and the constants of \eqref{FQcon} and \eqref{GQcon} expressed in terms of Fourier series are
\begin{eqnarray}
\mathcal{H}&=& \frac{1}{2} \sum_{\mathbf{k}} \bigg((\rho_s^2+k^{-2})|\omega_\mathbf{k}|^2 
+ \frac{k^2 |\psi^\star_\mathbf{k}|^2}{1+k^2 d_e^2} \bigg)\,,
\label{Heqfourier}\\
F &=&  \frac{1}{2} \sum_\mathbf{k} ( |\psi^\star_\mathbf{k}|^2 + d_e^2\rho_s^2  |\omega_\mathbf{k}|^2)\,,
\label{Feqfourier}\\
G &=&  \frac{1}{2}\sum_\mathbf{k} \big( \omega_\mathbf{k}\psi^\star_\mathbf{-k} + \omega_\mathbf{-k}\psi^\star_\mathbf{k} \big)\,.
\label{Geqfourier}
\end{eqnarray}
Equations \eqref{Feqfourier} and  \eqref{Geqfourier} can be thought of as 2D remnants of the magnetic and cross helicities\cite{lingam16} if we set $\rho_s = 0$,  although since there is no third dimension they lose their topological meaning associated with linking.  It can be shown via direct calculation that  these  helicity remnants are indeed rugged.  For instance, using \eqref{fourierpsieq} and  \eqref{fourieromegaeq} and the reality condition $\overline{\omega}_{\mathbf{k}} = \omega_{-\mathbf{k}}$ with overbar being complex conjugate,  we find
\begin{equation}
	\dot G =\!\! \sum_{\mathbf{k},\mathbf{k}^{\prime\prime}} \hat z \cdot \mathbf{k}^{\prime\prime}\times \mathbf{k}\, \lf((\mathbf{k}+\mathbf{k}^{\prime\prime})^2\,\psi^\star_{\mathbf{k}} \psi^\star_{\mathbf{k}^{\prime\prime}}  +\rho_s^2 \omega_{\mathbf{k}} \omega_{\mathbf{k}^{\prime\prime}} \ri) \psi_{\mathbf{k}^\prime}= 0\,.
\end{equation}
Similarly, it is not hard to show that a Liouville theorem is satisfied, i.e., 
\begin{equation}
	\f{\p \dot \omega_{\mathbf{k}}}{\p\omega_{\mathbf{k}}} = 0
 \quad\text{and}\quad \f{\p \dot \psi^\star_{\mathbf{k}}}{\p\psi^\star_{\mathbf{k}}} = 0\,.
 \label{liou}
\end{equation}
It is necessary to demonstrate this in order to apply equilibrium statistical mechanics,  even though the  model of \eqref{2dxmhdrealequations} is Hamiltonian. This is because the variables $\psi^\star$ and $\omega$ are noncanonical and one must identify an invariant measure.  The Darboux theorem assures that the usual phase space volume  measure is preserved in some local canonical coordinate system;\cite{morrison98} however, in the truncated noncanonical coordinates, the finite number of retained Fourier amplitudes, equations  \eqref{liou} need to be verified.  We emphasize this point because sometimes this step is missing in analyses. 

\section{Absolute Equilibrium States}
\label{AES}

\begin{figure}
	\begin{tikzpicture}[scale=.5]
	\foreach \x in {-6,-4,...,6}	
	\draw[red,<-,very thick] (\x,0) arc (0:180:1);
	\draw[->,very thick]  (-8.5,-.1) -- (7,-.1) node[very near start,anchor=north] {Driving Range} node[midway,anchor=north] {Inertial Range} node[very near end,anchor=north] {Dissipative Range} node[anchor=south] {$k$};
	\end{tikzpicture}
	\caption{\small Schematic  demonstrating the  standard Richardson-Kolmogorov direct cascade. Energy injected at  low $k$,  e.g.,  via large scale stirring, cascades through the inertial range and dissipates at small scales (large $k$). Upon reversal of the arrows along with the driving and dissipative ranges, a depiction of the inverse cascade is obtained.}\label{cascadecartoon}
\end{figure}
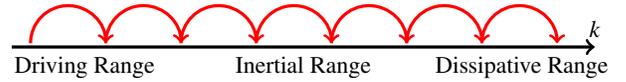

{We now turn to our study  of turbulent cascades using the statistical mechanics  of AES,  even though turbulence is an out-of-equilibrium phenomenon. This might be seen as counterintuitive; however,  it important to stress here that the AES hypothesis is a tool used to predict the direction of cascades \cite{frisch95,biskamp03}and does not in general describe the distribution of actual invariants in fully developed turbulence in a    driven dissipative system.  The operative intuitive idea is that the AES captures the relevant properties of the  nonlinear dynamics active in the inertial range. {For instance, in 2D HD turbulence, using AES one could infer the presence of the inverse energy cascade that dumps energy to large scales away from the small scales where the dissipation normally occurs.}  As a consequence, the flow dynamics is dominated by large scale coherent structures, such as vortices or jets.\cite{bouchet12}} In the 3D fluid case there is the  well-established cascade\cite{kolmogorov41} (see Fig.~\ref{cascadecartoon}) from large scales, where  stirring occurs,  to the short scales,  where  energy is dissipated, a picture that has been confirmed in experiments.\cite{frisch95,sreenivasan97,biskamp03}   So typically one follows Kolmogorov and makes use of phenomenological {estimates based} on dimensional  {arguments} in order to describe  turbulent spectra. For instance this was done in  recent work on a 3D extended MHD model,  where steepening of spectra were predicted\cite{abdelhamid16}  and in a companion work\cite{miloshevich17}   the direction of such cascades were investigated. This work relied on a generalization of  pioneering works in hydrodynamics\cite{lee52,kraichnan67} and MHD turbulence\cite{frisch75,lee52} based on statistical mechanics ideas.  We  apply those same methods here.

The idea\cite{lee52,burgers29a} is to assume that  Fourier modes play a role analogous to that  of the  particle degrees of freedom in statistical mechanics. One   calculates spectra in  the canonical ensemble,  and then makes predictions regarding the direction of the cascades based on where the spectra peak. It is understood that in reality  dissipation acts to remove the ultraviolet catastrophe (high $k$ divergence) that typically occurs in Galerkin systems.\cite{kraichnan80}  

There is a problem that may arise in a case when one has non-additive constants of motion that may lead to non-Boltzmann statistics. For more on this see the discussion in {Ref.}~\onlinecite{isichenko94}. 
On the other hand, in the case of the 2D Euler equation, we find that according to {Ref.}~\onlinecite{campa09}, even though the canonical distribution has to be used with caution for long-range interacting systems, the statistical tendency of vortices of the same sign of   circulation  to cluster in the so-called negative temperature regime can be indeed predicted using the same canonical distribution by observing that spectra peak at low $k$.

\begin{figure}
	\begin{center}
		\subfloat[Total Energy $H$]{\includegraphics[width=.35\textwidth]{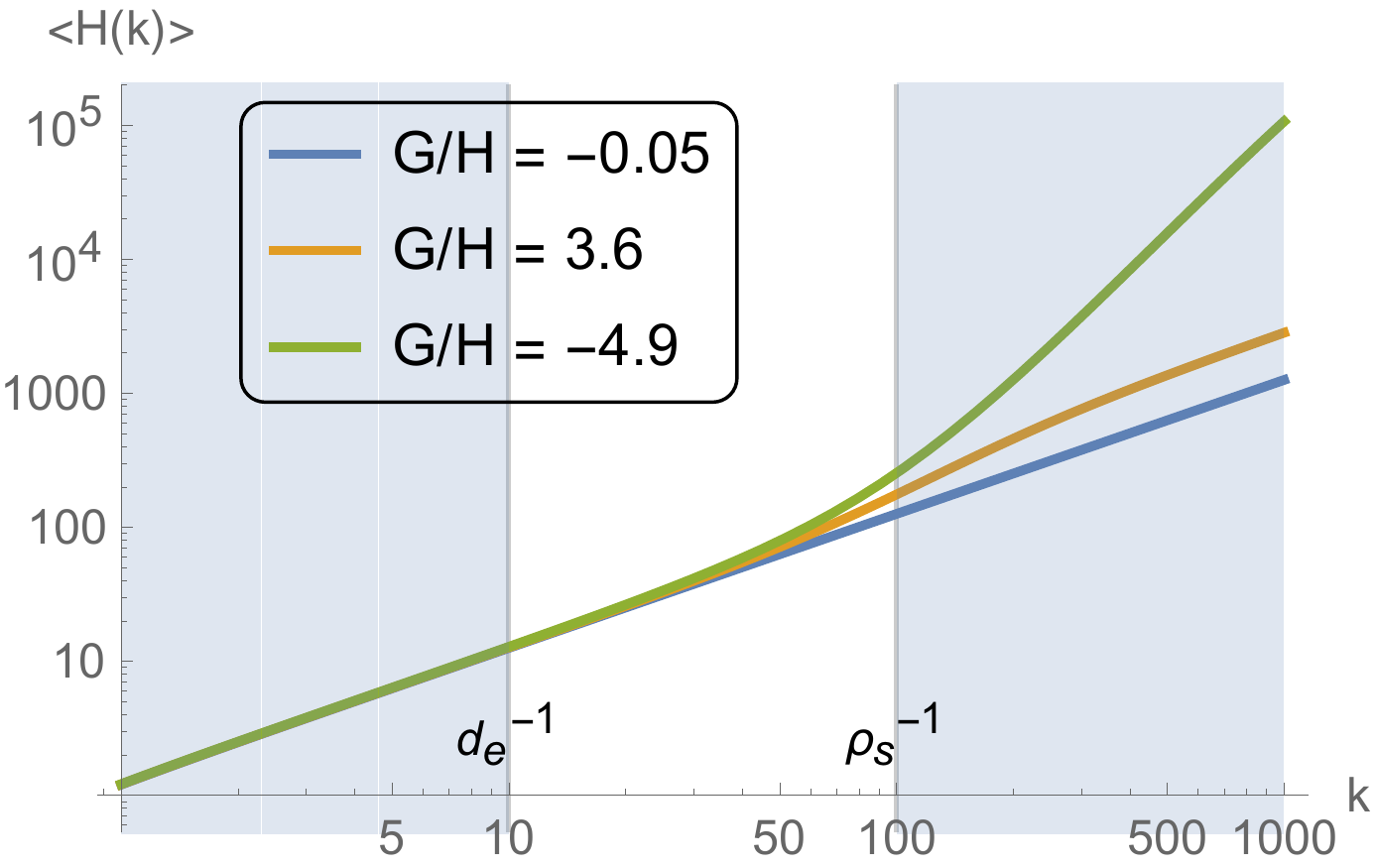}\label{dehamiltonian}}\\
		\subfloat[Remnant Magnetic Helicity $F$]{\includegraphics[width=.35\textwidth]{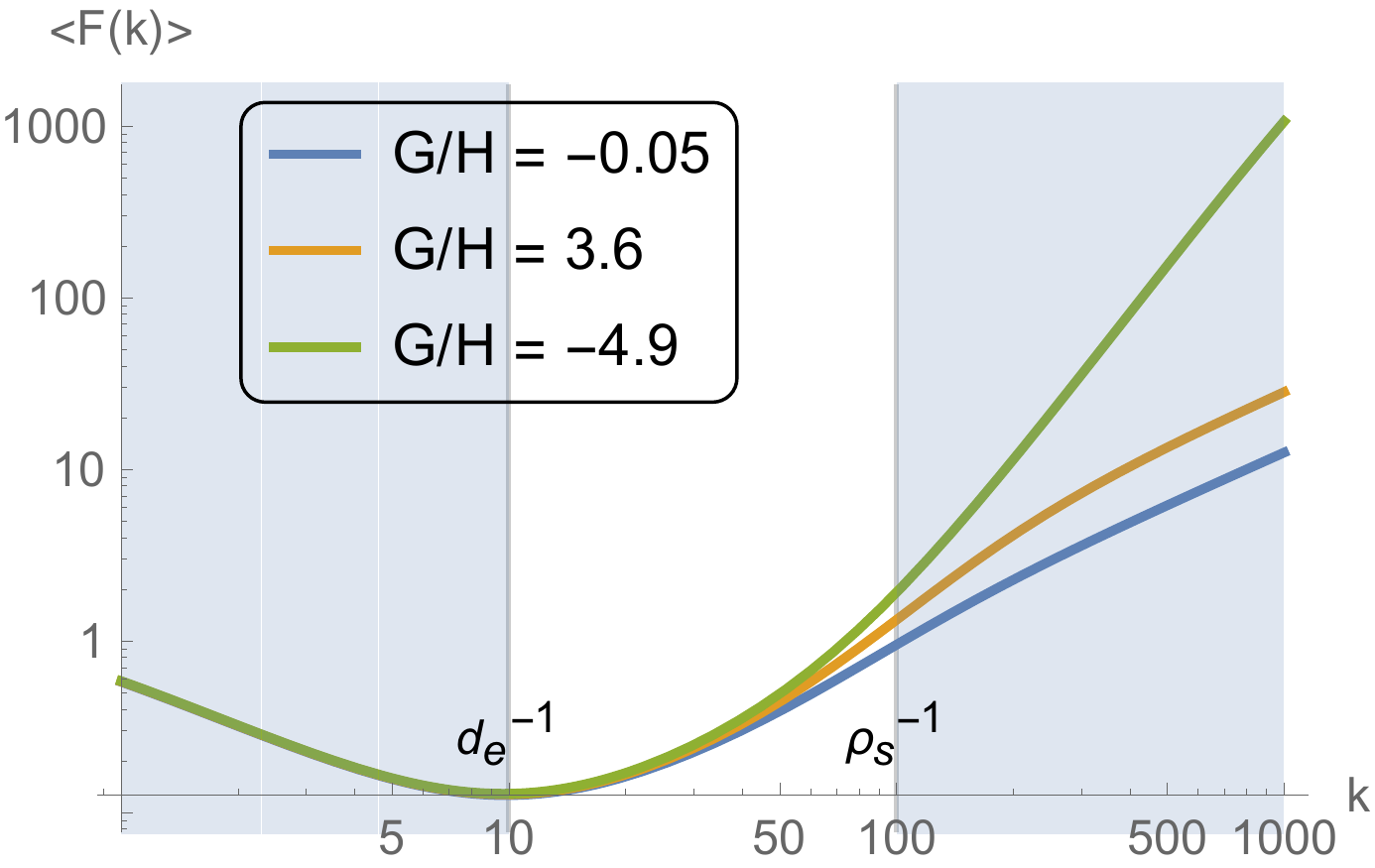}\label{demagF}}\\
		\subfloat[Remnant Cross Helicity $G$]{\includegraphics[width=.35\textwidth]{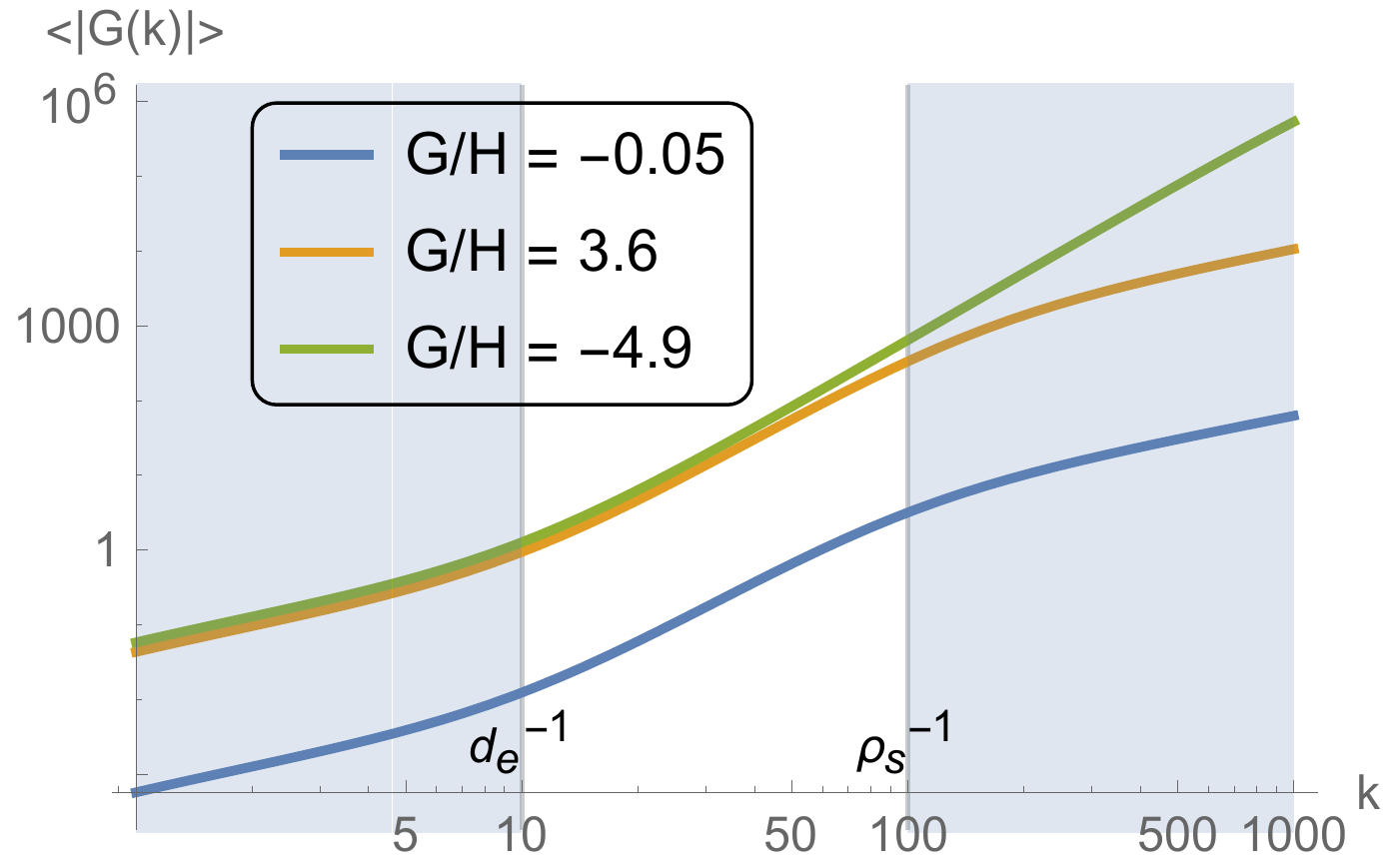}\label{decross}}
	\end{center}
	\caption{Log-Log Plots for total energy, remnant magnetic helicity {and cross-helicity}. The parameters used here are $\alpha = 10$, $\beta = 1$ and 
$\gamma = \{0,-.75,1\}$ is varied so that different values of G are obtained (color-coded, see the legend for the description). The microscales were chosen to be $d_e = 0.1$ and $\rho_s = 0.01$. As a result there are no constraints imposed on $k$ according to \eqref{normalizability}. The main feature is the change of the sign of the slope of $F = F(k)$ as we  smoothly transition through the $k \sim d_e^{-1}$ scale. Notice that plots are obtained under the assumption that $k_\text{min} = 1$ and not $2\pi$ for the simplicity.}
\label{deenergy}
\end{figure}

We seek AES given by the phase space probability density of the form
\begin{equation}
\mathcal{P} = Z^{-1} e^{-\alpha \mathcal{H} - \beta F - \gamma G} =: Z^{-1} e^{- A^{ij}u_i u_j/2}\,,
\label{Z}
\end{equation}
where $u_i := \{\omega^{\Re}_\mathbf{k}, \omega^{\Im}_\mathbf{k},\psi^{\star\Re}_\mathbf{k},\psi^{\star\Im}_\mathbf{k}\}$ and according to \eqref{Heqfourier}, \eqref{Feqfourier}, and \eqref{Geqfourier}  the  matrix $(A^{ij})$ is given by
\begin{equation*}
A:=\begin{pmatrix}
\delta & 0 & \gamma & 0 \\
0 & \delta & 0 & \gamma \\
\gamma & 0 & \eta & 0 \\
0 & \gamma & 0 & \eta
\end{pmatrix},
\end{equation*}
where 
\begin{equation}
\delta := (\alpha + \beta d_e^2)\rho_s^2 + \dfrac{\alpha}{k^2}\quad \mathrm{and}\quad
\eta := \dfrac{\alpha k^2}{1+k^2 d_e^2}+\beta\,.
\end{equation}
{The parameters $\alpha$, $\beta$ and $\gamma$ present in Eq. \eqref{Z} are Lagrange multipliers. Their values in terms of the parameters $d_e$ and $\rho_s$ are determined by a normalization condition and  by imposing that the expectation values of the invariants $\mathcal{H}$, $F$ and $G$  match their initial values  (see Eqs. (\ref{expectationG}), (\ref{Fexpectation}) and (\ref{hamexpectation})  ). This will be carried out in Sec. \ref{sec:regimes}. These Lagrange multipliers are akin to the  inverse temperatures found in statistical mechanics. }

Using \eqref{Z} the partition function $Z$ follows from the  normalization condition
\begin{equation}
	\int \calP(k)\, d\Gamma(k) = \int \calP(k) \;d\psi^{\star\Re}_\mathbf{k} \,d\psi^{\star\Im}_\mathbf{k}
	 \,d\omega^{\Re}_\mathbf{k} \,d\omega^{\Im}_\mathbf{k}= 1,
\end{equation}
where $\psi^{\star}_\mathbf{k} =: \psi^{\star\Re}_\mathbf{k} + i\, \psi^{\star\Im}_\mathbf{k}$. 
Because the statistics are Gaussian, integration is straightforward and  the partition function is found to be  
\begin{equation}
	Z = \f{(2\pi)^2}{\sqrt{\det{A}}}\,.
\end{equation}

\begin{figure}
	\begin{center}
		\subfloat[Total Energy $H$]{\includegraphics[width=.35\textwidth]{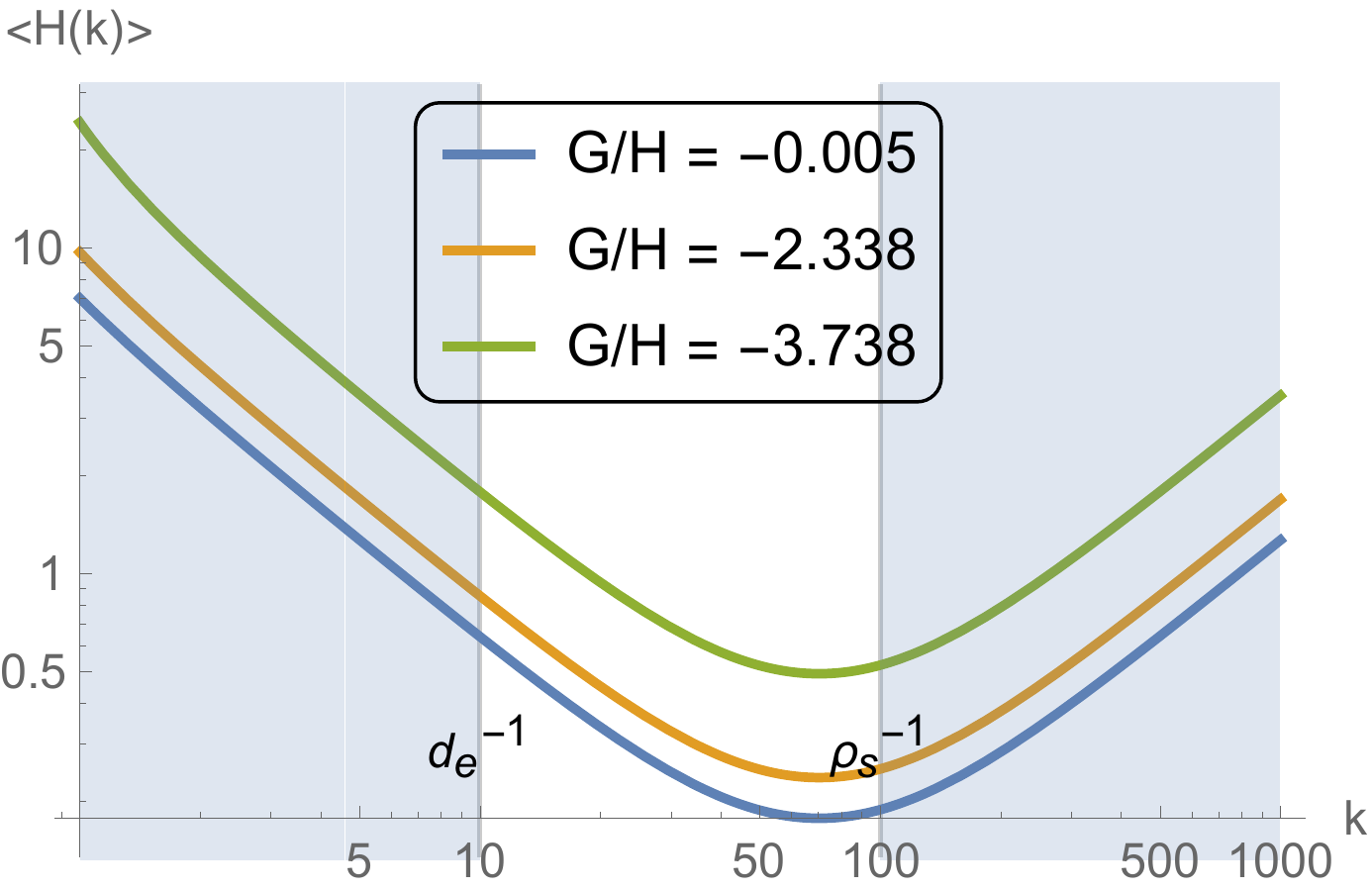}\label{gyroonlyenergy}}\\
		\subfloat[Remnant Magnetic helicity $F$]{\includegraphics[width=.35\textwidth]{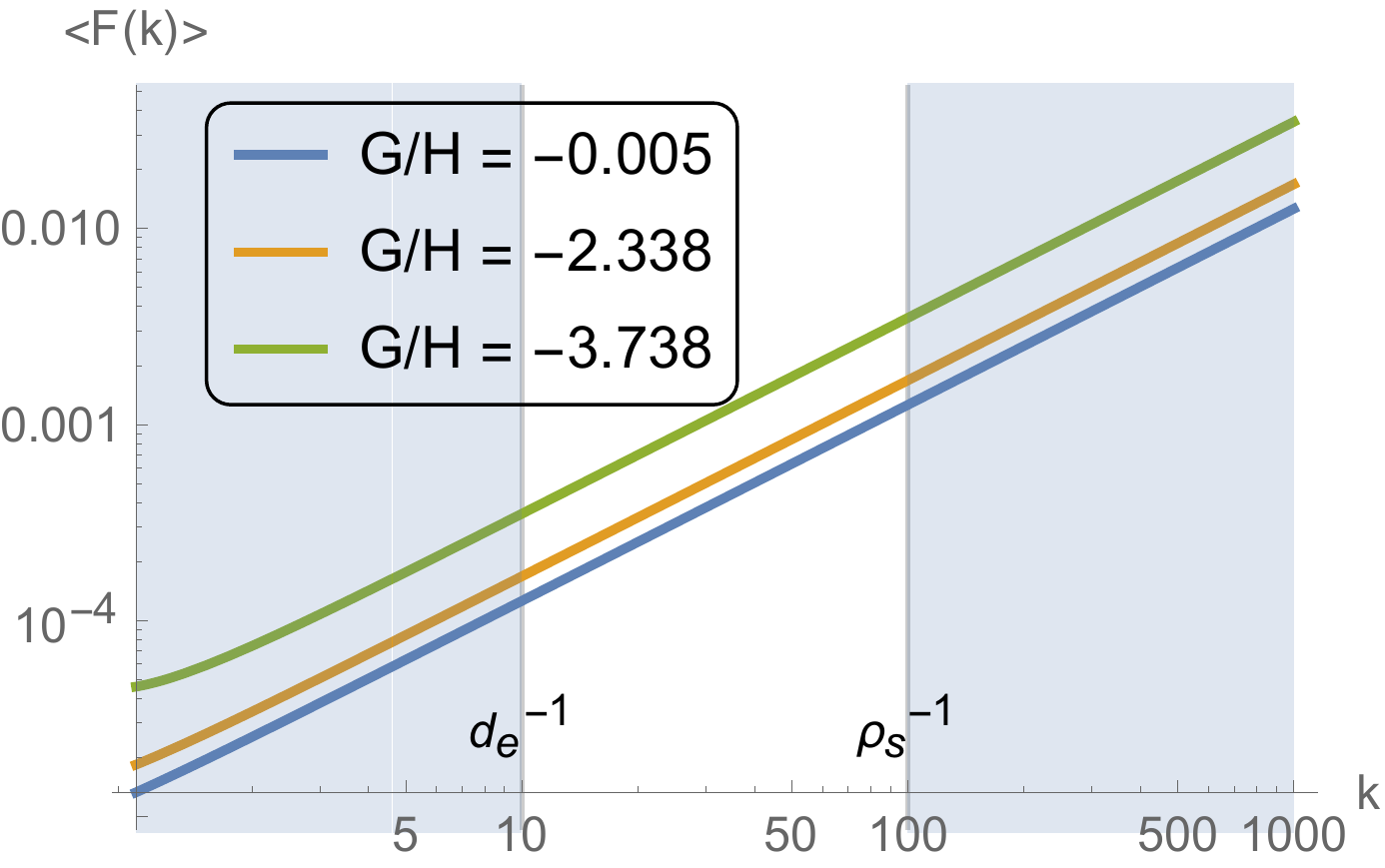}}\\
		\subfloat[Remnant Cross Helicity $G$]{\includegraphics[width=.35\textwidth]{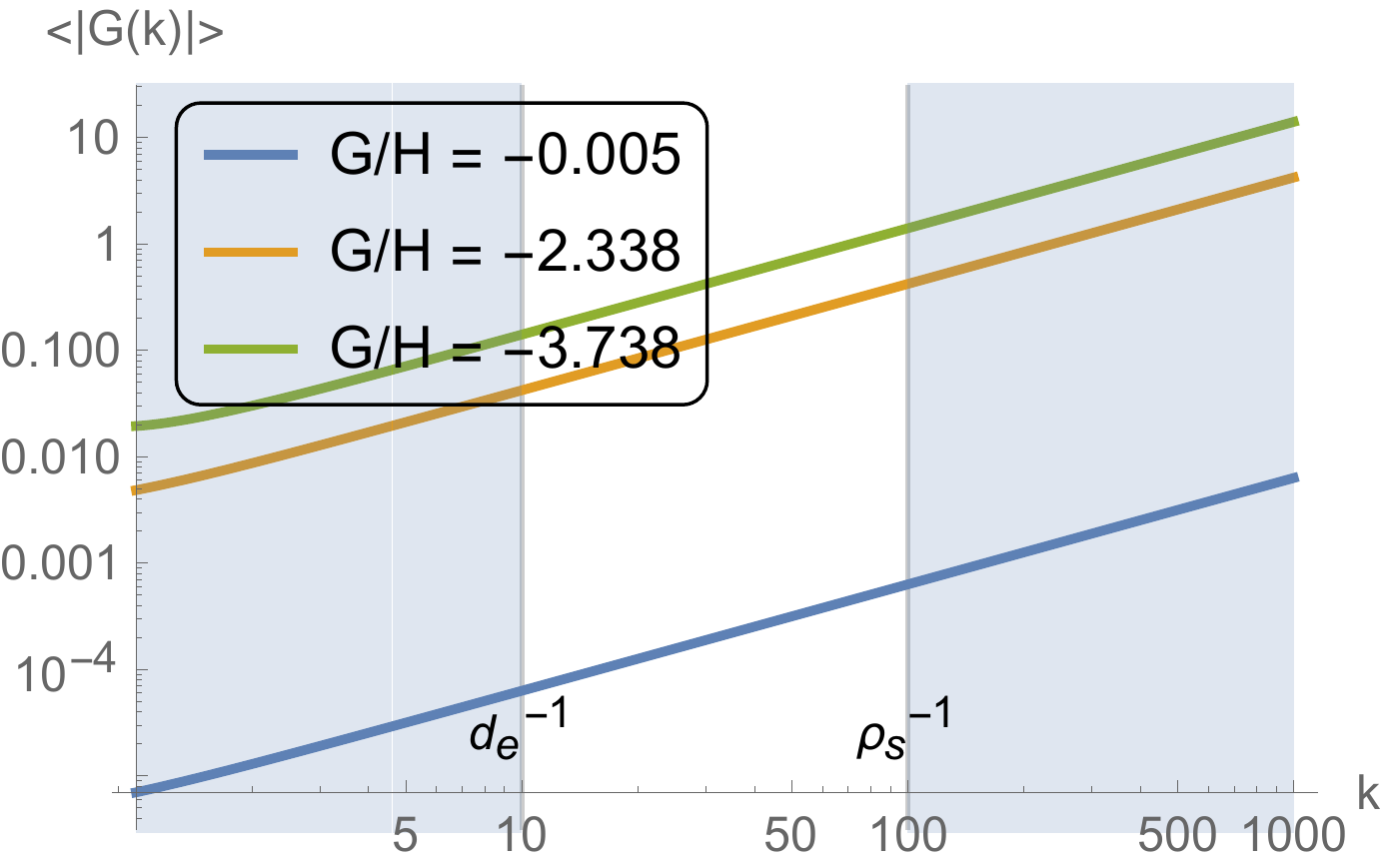}\label{crosshelicitygyro}}
	\end{center}
	\caption{Log-Log Plots for total energy, {remnant magnetic helicity and cross-helicity}. {The parameters used here are $\alpha = -0.1$, $\beta = 10^6$ and $\gamma = \{0,500,800\}$ is varied so that different values of $G$ are obtained (color-coded, see the legend for the description). The microscales were chosen to be $d_e = 0.1$ and $\rho_s = 0.01$. Helicity $F$ seems to only have direct cascade, when $\alpha < 0$. This can also be seen from Table \ref{HandFacros} since $\beta$ is so large. The highlight of these negative energy states is the possibility of the inverse cascade of energy that seems to be independent of the cross-correlation $\gamma$. }}\label{gyroenergy}
\end{figure}

One can also invert the matrix $A$  to obtain various expectation values, such as   $\langle u_i u_j \rangle = A^{-1}_{ij}$.\cite{kardar07}  In addition, it is necessary to investigate the realizability condition that  the matrix $A$ needs to be positive-definite. Thus we impose the condition of positivity of its eigenvaules, otherwise the probability distribution would not be  integrable. After some algebra we arrive at the following inequalities
\begin{eqnarray}
\label{normalizability}
(\alpha+\beta d_e^2)\rho_s^2 k^2  + \alpha &>& 0\,,\label{normfirst}
\\ 
(\alpha + \beta d_e^2) k^2  + \beta &>& 0  \label{normsecond}
\\
 \Big\lbrack (\alpha+\beta d_e^2)\rho_s^2 k^2 + \alpha \Big\rbrack \Big\lbrack (\alpha &+& \beta d_e^2) k^2  + \beta  \Big \rbrack 
\nonumber\\
&>& k^2 (1+k^2 d_e^2)\gamma^2\,. 
\end{eqnarray}
At this point it is important to observe that $\alpha > 0$ when we set $\rho_s = 0$. Thus {2D planar incompressible XMHD} cannot have the so-called ``negative temperature states'' (NTS) that correspond to $\alpha < 0$. It appears that NTS are in principle possible if $\rho_s$ is not ignorable,  {i.e., when thermal electron effects are taken into account}. This is interesting since it is known that in 2D fluid turbulence they are associated with the inverse cascade of energy.\cite{onsager49}  Actually NTS have been analytically predicted in gyrokinetics \cite{zhu10} in the 2+1D case as well as in some earlier works on drift-wave turbulence.\cite{gang89,koniges91}  {The latter works consider  fluid models formed by an incompressible Euler equation together with an equation for an advected scalar. Therefore they differ qualitatively from the model  \eqref{2dxmhdrealequations}} that we are using . 

It is evident from \eqref{normfirst} that if $\alpha < 0$ then $\tilde{\alpha}:=\alpha+\beta d_e^2 > 0$. Alternatively,  we can have $\alpha > 0$, which if $\beta > 0$ obviously implies $\tilde{\alpha}>0$ and on the other hand if $\beta < 0$ then \eqref{normsecond} implies that $\tilde{\alpha}$ is again positive. Thus we have the useful inequality independent of $k$
	\begin{equation}\label{alphatilde}
		\tilde{\alpha}:=\alpha+\beta d_e^2 > 0\,.
	\end{equation}

We proceed with evaluating various expectations of correlations. The quantities of interest are the average {squared} generalized flux function per wave-mode
\begin{equation}
\frac{1}{2}\langle |\psi^*_\mathbf{k}|^2 \rangle  = \Big \lbrack \frac{{\tilde{\alpha}}k^2 + \beta}{1+k^2 d_e^2} -\gamma^2 \frac{1}{{\tilde{\alpha}}\rho_s^2 + \alpha k^{-2}} \Big\rbrack ^{-1}
\end{equation}
{ and the} average {squared} vorticity
\begin{equation}
\frac{1}{2} \langle |\omega_\mathbf{k}|^2\rangle = \Big\lbrack {\tilde{\alpha}} \rho_s^2 + \frac{\alpha}{k^2}- \gamma^2\frac{1+k^2 d_e^2}{{\tilde{\alpha}}k^2 + \beta}\Big\rbrack^{-1}.
\end{equation}
To calculate the remnant cross-helicity we need to add cross-correlation terms
\begin{equation} \label{expectationG}
\langle G (k) \rangle = {-\f{\gamma}{\lf(\tilde{\alpha} \rho_s^2+\dfrac{\alpha}{k^2}\ri)\dfrac{\tilde{\alpha} k^2 + \beta}{1+k^2 d_e^2} - \gamma^2}\,.}
\end{equation}
To simplify the analysis we assume  that  the remnant {cross-helicity $G$ is zero} and therefore $\gamma = 0$. Thus,  per wave mode,  we obtain the expressions 
\begin{eqnarray}
\label{Fexpectation}
\langle F (k) \rangle &=&  \f{d_e^2 \rho_s^2 k^2}{\alpha + \tilde{\alpha}\rho_s^2 k^2}+ \f{1+k^2 d_e^2}{\tilde{\alpha} k^2 + \beta}\,,
\\
\label{hamexpectation}
\langle H (k)\rangle &=& \f{1+\rho_s^2 k^2}{\alpha + \tilde{\alpha} \rho_s^2 k^2} + \f{k^2}{\tilde{\alpha} k^2 + \beta}.
\end{eqnarray}
{This is consistent with the MHD results of {Ref.}~\onlinecite{fyfe76} (if we relabel appropriately $\alpha\rightarrow2\alpha, \beta\rightarrow2\gamma,\gamma\rightarrow2\beta$ and set $\rho_s \rightarrow 0$ and $d_e \rightarrow 0$). }

{We observe that, for large $k$, the remnant helicity and energy spectra  behave as follows:}
 \begin{equation}\label{FHMHDbeh}
2\pi k \langle F(k) \rangle \approx \mathcal{O}\Big({1}/{k}\Big), \quad 2\pi k \langle \mathcal{H} (k)\rangle \approx \mathcal{O}(k)\,, 
\end{equation}
{similarly to MHD. On the other hand, at large scales, the presence of finite electron temperature
can yield a different behavior, depending on the value of parameters. Relevant limits of the remnant
helicity and energy spectra will be discussed in Sec.~\ref{comparisons}.}

%
%
%
%

 \section{Qualitative analysis}  \label{sec:regimes}

In this section we will discuss different regimes that the system exhibits. The parameters $\alpha$ and $\beta$ can be found from the total energy and the remnant helicity,  which are obtained as $H = \int 2\pi k \,\langle H(k) \rangle\, dk$ {and $F = \int 2\pi k \,\langle F(k) \rangle\, dk$. It also turns out to be convenient to introduce the variable} $\tilde{F} := F - d_e^2 H$ and the ratio 
	\begin{equation}
	\label{kratio}
		K := \f{H}{\tilde{F}} = 2\f{k_\text{max}^2 - k_\text{min}^2}{\ln{\dfrac{(\beta + \tilde{\alpha}k_\text{max}^2)(\alpha + \tilde{\alpha}\rho_s^2k_\text{min}^2)^{d_e^2/\rho_s^2}}{(\beta + \tilde{\alpha} k_\text{min}^2)(\alpha + \tilde{\alpha}\rho_s^2 k_\text{max}^2)^{d_e^2/\rho_s^2}}}} - \f{\beta}{\tilde{\alpha}}\,.
	\end{equation}
	Notice that $\alpha H + \beta F = \tilde{\alpha}H + \beta \tilde{F}$; however,  since $\tilde{F}$ is not a Casimir, { in the following we will focus on the invariant $F$}. In addition we observe the well-known identity
	\begin{equation}
		\alpha H + \beta F = 2\pi\,(k_\text{max}^2 - k_\text{min}^2)\,.
	\end{equation}
	For simplicity we first consider the  {2D planar incompressible XMHD} limit $\rho_s\rightarrow 0$. Then \eqref{kratio} becomes 
	\begin{equation}\label{kratio2}
	K \rightarrow \f{2}{(k_\text{max}^2 - k_\text{min}^2)^{-1}\ln{\dfrac{\beta + \tilde{\alpha}k_\text{max}^2}{\beta + \tilde{\alpha} k_\text{min}^2}} -d_e^2\dfrac{\tilde{\alpha}}{\alpha}} - \f{\beta}{\tilde{\alpha}}\,.
	\end{equation}
	
\begin{figure}
\begin{tikzpicture}
\draw[->] (-3,0) -- (4.2,0) node[right] {$\dfrac{\beta}{\alpha}$};
\draw[->] (0,-3) -- (0,3) node[above] {$K$};
 \draw [cyan,thick] (-0.5,.5) .. controls (-0.5,2) and (1.9,1.5) ..(1.9,3);
 \draw [cyan,thick] (2.1,-3) .. controls (2.1,-2) and (2,-1.1) .. node[very near end,sloped,below] {XMHD} (4,-1.1);
 \draw[red,dashed] (2,-3) -- (2,3);
 \draw[green!50!black,dashed] (0,-1) -- node[near start,sloped,below] {$K_\infty = -d_e^{-2}$} (4,-1);
 \draw[green!50!black,dashed] (-.5,.5) -- node[very near end,sloped,right] {$K_\text{cr} = k^2_\text{min}$} (0,.5);
 \filldraw [gray] (-0.5,.5) circle (1pt);
  \draw [cyan,thick] (-1.5,1) .. controls (-1.5,0) and (-2,-1) .. node[near end,sloped,below] {forbidden} (-3,-1);
   \filldraw [gray] (-1.5,1) circle (1pt);
   \fill[pattern=north west lines, pattern color=green] (-3,-3) rectangle (-1,3);
  \draw [purple,thick] (-0.6,.5) .. controls (-0.5,2) and (1.5,2) .. node[very near end,sloped,below] {MHD} (4,2); 
  \filldraw [gray] (-0.6,.5) circle (1pt);
  \filldraw [green!50!black] (0,1.5) circle  (1pt) node[above left]{$K_b$};
\end{tikzpicture}
\caption{Description of the $K$ vs $\beta/\alpha$ dependence (not to scale) according to \eqref{kratio2} when $\rho_s = 0$. }\label{schematicsofk}
\end{figure}
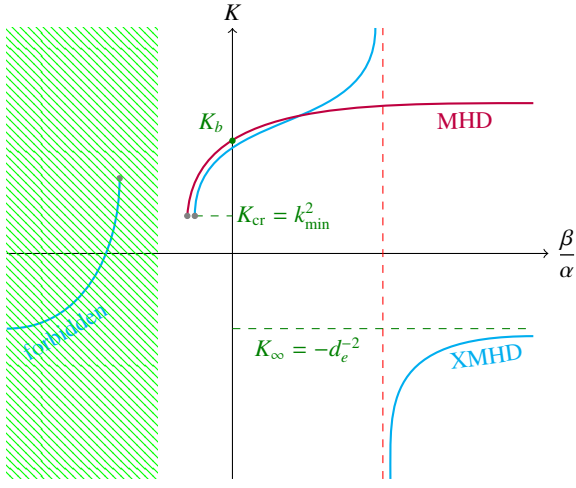

The parameter $\beta$ switches sign at
	\begin{equation}  \label{Kbswitch}
	K_b := K(\beta = 0) = \f{2}{(k_\text{max}^2 - k_\text{min}^2)^{-1} \ln \dfrac{k_\text{max}^2}{k_\text{min}^2} - d_e^2}\,,
	\end{equation}
{signaling the emergence of negative temperature states.}  Notice that $K_b > 0$ provided $d_e$ is small {enough}. The local minimum is reached when
\begin{equation} \label{Kcritical}
	K_\text{cr}:=K\lf(\f{\beta}{\alpha} = -\f{k^2_\text{min}}{1+k^2_\text{min}d_e^2} \ri) = k^2_\text{min}\,.
\end{equation}
A depiction of the  behavior is  shown in Fig.~\ref{schematicsofk}. Notice that at $K_\text{cr}$ the remnant helicity condenses to the lowest wavenumber $k_\text{min}$. This can be seen from the second term in \eqref{Fexpectation} and is a direct analogy of the energy condensation in HD proposed by Kraichnan \cite{kraichnan75} and others.

In addition, it can be shown that the logarithm found in the denominator of \eqref{kratio2} is a monotonically decreasing function of  $\beta/\alpha$  because $k_\text{max} > k_\text{min}$, while the magnitude of the second term is linearly increasing and thus there exists a pole. This pole is absent in MHD, where therefore $K > 0$. This will be important below. The analysis is concluded by observing that as $\beta/\alpha\rightarrow \infty$,  $K$ approaches $K_\infty = -d_e^{-2}$ and thus {curiously there seems to be a gap in the admissible values of $K$}.

Now let us step back to MHD by letting $d_e\rightarrow 0$ and explicitly follow an argument found in {Refs.}~\onlinecite{fyfe76,fyfe77}. In this case the following identity can be found from \eqref{kratio2} in the limit $k_\text{max} \rightarrow \infty$:
\begin{equation}
	\f{\beta}{\alpha} + k_\text{min}^2 = k_\text{max}^2 \exp\lf[{-\f{2 k_{\max}^2}{K}}\ri]\rightarrow 0\,.
\end{equation}
Thus,  the authors conclude that physically one can expect condensation to the lowest wavenumber since $\beta$ becomes negative. {If $\beta$ is negative we can have a low-lying pole as will be described below. And when $K$ reaches it local minimal value (associated with a specific negative value of $\beta$, see 
Fig.~\ref{schematicsofk} and Eq.~\eqref{Kcritical}) then this pole coincides with $k_\text{min}$. Existence of a pole naturally implies that most of the spectral quantity is going to condense there. }

If we redo these arguments for the  XMHD case,  we obtain
\begin{equation}
	K\rightarrow - d_e^{-2}\lf(\f{\alpha}{\tilde{\alpha}} + 1\ri)\Rightarrow \f{\beta}{\alpha} \rightarrow - d_e^{-2}\lf (1+d_e^2 K\ri)
\end{equation}
and therefore $\beta/\alpha$ may remain positive,  thus avoiding condensation for some values of $K$ even if $k_\text{max}\rightarrow\infty$. 

When the electron temperature is not ignorable ($\rho_s > 0$) we recover the $\alpha < 0$ regime and the situation becomes more complicated according to \eqref{kratio}. From \eqref{Fexpectation} and \eqref{hamexpectation} we can see that there are two poles. In the vicinity of one pole the other term can be ignored. When $\beta$ is negative the remnant helicity condenses to $k_\text{cr,1}\sim \sqrt{-\beta/\tilde{\alpha}}$ as described above in the XMHD case. However in the $\alpha < 0$ case the pole $k_\text{cr,2}\sim \sqrt{-\alpha\rho_s^{-2}/\tilde{\alpha}}$ dominates and the roles of $H$ and $F$ are interchanged. Notice that both poles cannot occur simultaneously since that would clearly violate \eqref{alphatilde}. When $\rho_s$ is small enough one expects a diagram similar to that of Fig.~\ref{schematicsofk}. It is not hard to show that $\beta$ changes sign at
\begin{equation}\label{K_bfull}
	K_b =2 \f{k_\text{max}^2 - k_\text{min}^2}{ \ln \dfrac{k_\text{max}^2}{k_\text{min}^2} - \dfrac{d_e^2}{\rho_s^2} \ln \dfrac{1+\rho_s^2 k_\text{max}^2}{1+\rho_s^2 k_\text{min}^2}}\,, 
\end{equation}
which generalizes {\eqref{Kbswitch}}. In fact, because the second term in the denominator is monotonic, it turns out that as a function of $\rho_s$ the quantity $K_b$ is bounded from below by {\eqref{Kbswitch}},  which is positive provided that $d_e$ is sufficiently small, so we can assume $K_b>0$. Similarly, $\alpha$ changes sign at
\begin{equation}
		K_a =-d_e^{-2}-\f{2\rho_s^2}{d_e^2} \f{k_\text{max}^2 - k_\text{min}^2}{ \ln \dfrac{k_\text{max}^2}{k_\text{min}^2} - \dfrac{\rho_s^2}{d_e^2} \ln \dfrac{1+d_e^2 k_\text{max}^2}{1+d_e^2 k_\text{min}^2}}
\end{equation}
and by the same argument $K_a < 0$,  provided that $\rho_s$ is sufficiently small.

\begin{figure}
	\begin{center}
		\subfloat[Total Energy $H$]{\includegraphics[width=.35\textwidth]{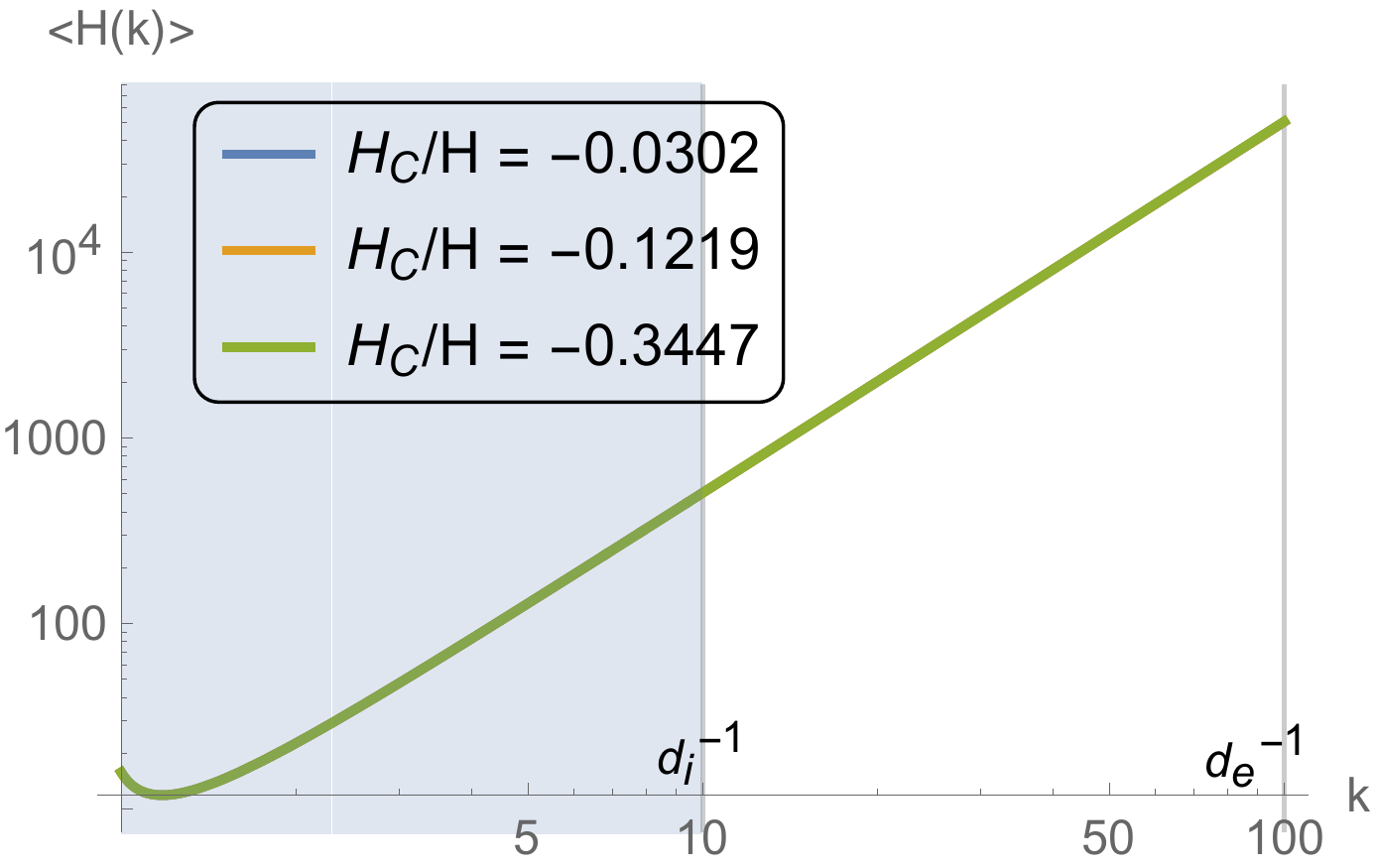}}\\
		\subfloat[Magnetic helicity $H_M$]{\includegraphics[width=.35\textwidth]{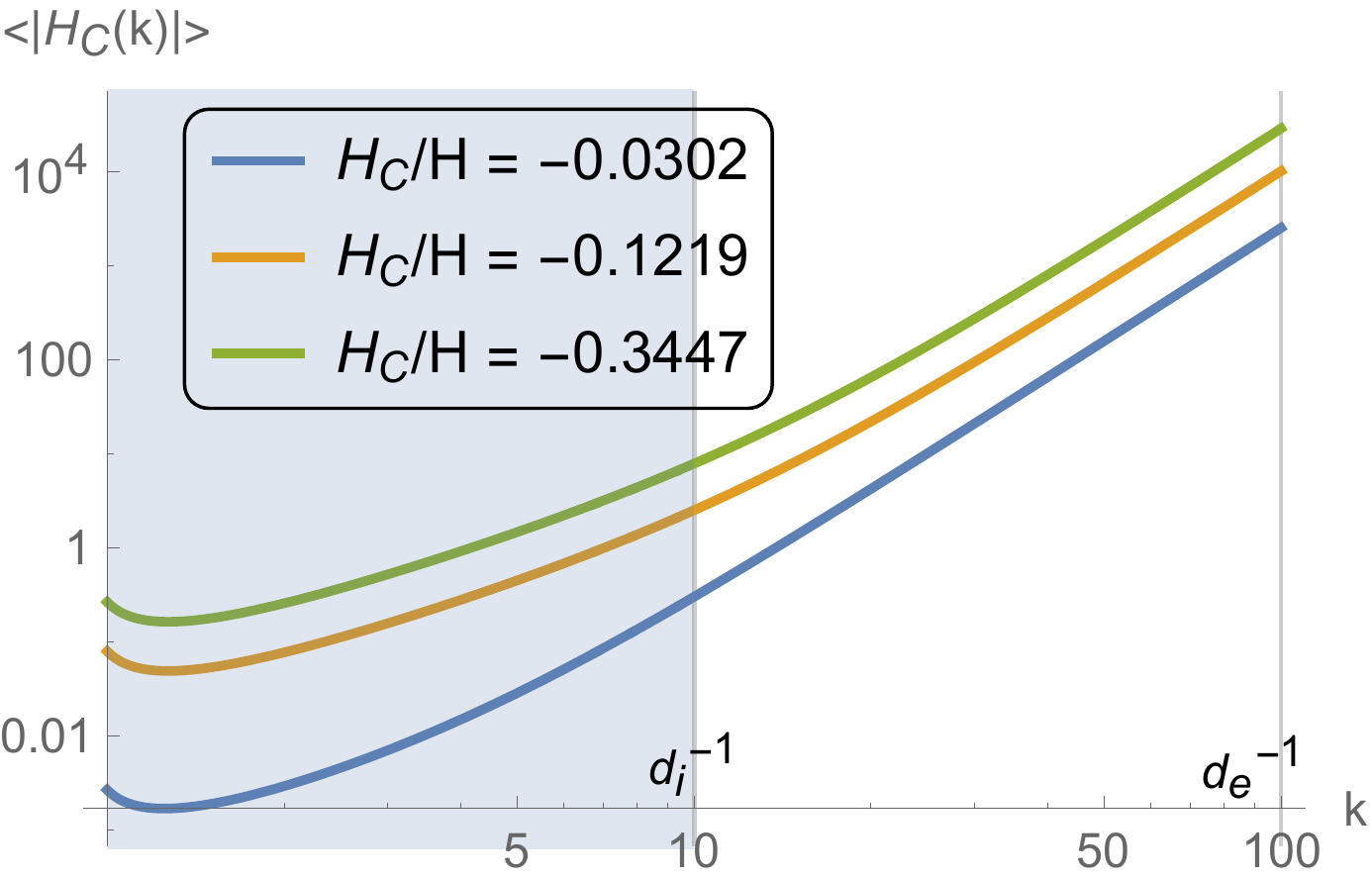}}\\
		\subfloat[Cross helicity $H_C$]{\includegraphics[width=.35\textwidth]{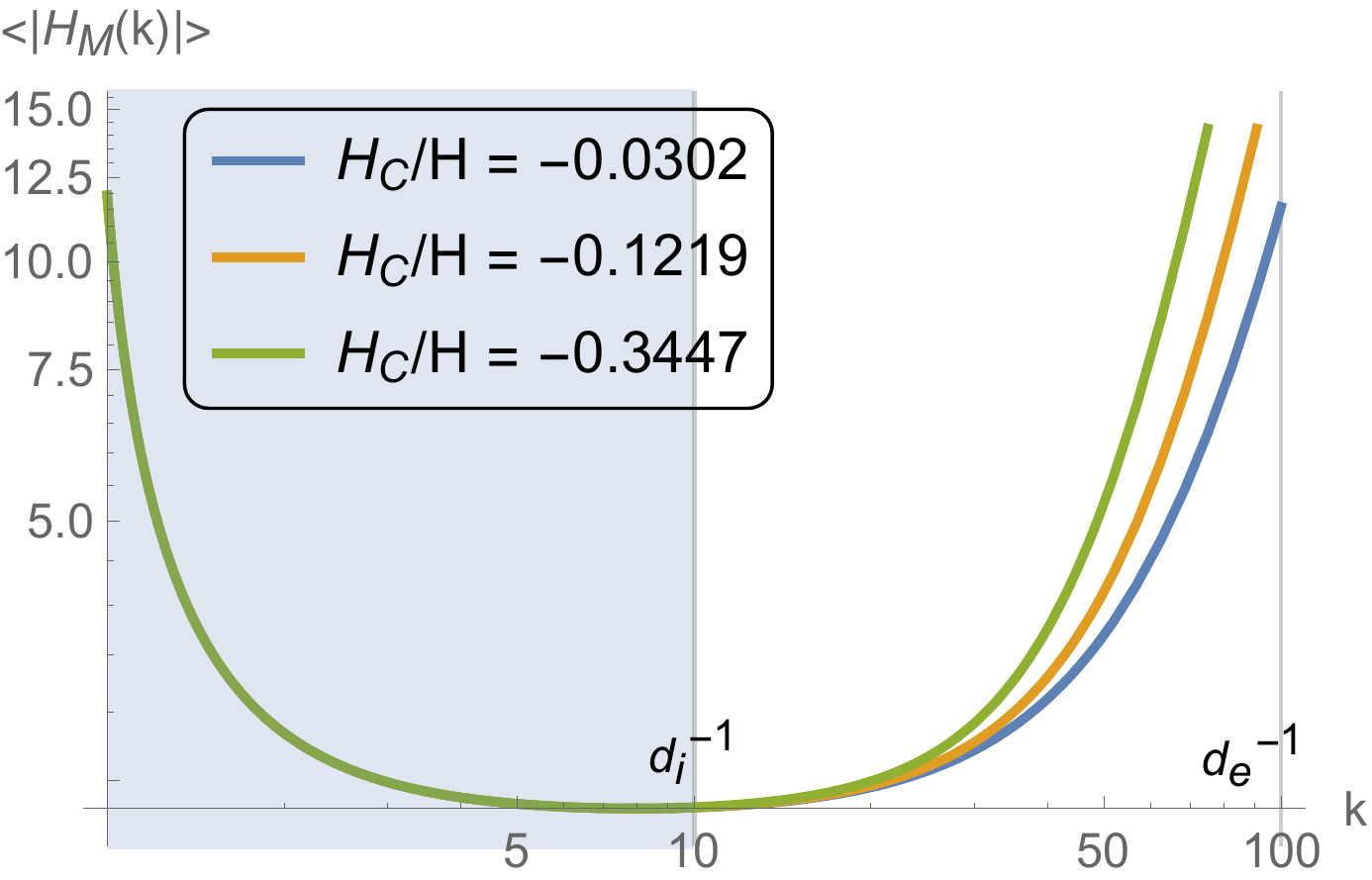}}
	\end{center}
	\caption{Log-Log Plots for total energy, magnetic and cross-helicity. {The parameters used here are $\alpha = 10$, $\beta = 9$ and $\gamma = \{0.001,0.03,0.1\}$ is varied so that different values of $H_C$ are obtained (color-coded, see the legend for the description). The microscales were chosen to be $d_i = 0.1$ and $d_e = 0.01$. }}\label{3dxmhdplots}
\end{figure}

\begin{table*}[htpb]
	\vspace{.25cm}
	\centering
	{\renewcommand{\arraystretch}{1.5}%
		{\setlength{\tabcolsep}{1.0em}
			\begin{tabular}{|r|c|c|}
				\hline
				Length Scale Choices& $\langle H (k)\rangle$ & $\langle F(k) \rangle$  \\
				\hline
				$1<k\ll(d_e^{-1},\rho_s^{-1})$ & $\dfrac{1}{\alpha}+\dfrac{1}{\alpha+\beta k^{-2}} $ & $\dfrac{1}{\alpha k^2 + \beta}$  \\[1ex]
				\hline
				$1\ll d_e^{-1}\ll k\ll\rho_s^{-1}$ & $\dfrac{1}{\alpha}+\dfrac{1}{\alpha+\beta d_e^{2}} $  & $\dfrac{1}{\alpha d_e^{-2} + \beta}$  \\[1ex]
				\hline
				$1\ll \rho_s^{-1}\ll k\ll d_e^{-1}$ & $\dfrac{1}{\alpha+\beta d_e^2}+\dfrac{1}{\alpha+\beta k^{-2}}$ & $\dfrac{1}{\alpha k^2 + \beta} + \dfrac{1}{\alpha d_e^{-2} + \beta}$ \\[1ex]
				\hline
				$1\ll (d_e^{-1},\rho_s^{-1})\ll k$ & $\dfrac{2}{\alpha+\beta d_e^2}$ & $\dfrac{2}{\alpha d_e^{-2} + \beta}$ \\[1ex]
				\hline
	\end{tabular}}}
	\caption{Various limits of spectral densities { when $\alpha > 0$ and $\beta$ not too large}. The first row corresponds to the large scale MHD limit; it was assumed that $\beta$ is not orders of magnitude larger than $\alpha$ to avoid singular perturbation and most likely this situtation is not realizable if one solves for the parameters via integrals of motion. The second row pertains to the  {2D planar incompressible XMHD} high $k$ limit, where gyroeffects have been ignored. The third row displays an opposite situation, where gyrophysics is relevant but the electron skin depth ignorable. The last row demonstrates the microscopic $k$ limit, and may be unphysical depending on how the model ordering works. Notice that terms were simply ignored based on the ordering, a more precise description would involve Taylor series.}
	\label{HandFacros}
\end{table*}
\section{Results, comparisons  and summary}
\label{comparisons}

Our  new results concern the limit $k d_e \gg 1$, where
\begin{equation}\label{FHXMHDbeh}
2\pi k \langle F(k) \rangle \approx \mathcal{O}(k), \quad 2\pi k \langle \mathcal{H}(k) \rangle \approx \mathcal{O}(k)\,.
\end{equation}
Thus we see that the scaling changes from the inverse to direct,  which suggests cascade reversal for the remnant magnetic  helicity $F$.
 Table \ref{HandFacros} contains our analyses for behavior across  scales when $\rho_s > 0$.  The cascade reversal behavior indicated by \eqref{FHMHDbeh} and \eqref{FHXMHDbeh} is seen in this more general analysis.  Thus there is  cascade reversal behavior at the electron skin depth in   {2D planar incompressible XMHD} as was predicted for  3D XMHD  in {Ref.~\onlinecite{miloshevich17}, although the details may vary.  In  Figs.~\ref{deenergy} and \ref{gyroenergy} we  plot spectral quantities for non-zero $\gamma$. 

As pointed out in {Ref.}~\onlinecite{fyfe76}, when $k d_e \ll 1$ the inverse cascade  {implies the presence of large scale structures in $\psi^\star$}. In our model due to the presence of $\rho_s$ this can be achieved in the  $\alpha>0$ regime; {on the other hand when $\rho_s = 0$ states $\alpha<0$ are forbidden}. {In 3D MHD,  the large scale presence of magnetic helicity is often associated with the generation of large scale magnetic fields.\cite{alexakis06,benerjee14}  In our previous work \cite{miloshevich17} we have explored the influence the electron inertia can have on the development of the turbulent cascade of the magnetic helicity in 3D. As stated earlier, in absence of $\rho_s$ the present  paper can be seen as a natural continuation of the earlier work, where geometry is simplified to two dimensions. In 2D MHD we see that instead of inverse cascade of magnetic helicity one has inverse cascade of the square vector potential and so we reach similar conclusions.} The fact that  magnetic helicity would condense to large scales is often cited as evidence of the dynamo action in MHD.\cite{fyfe77,biskamp90} The antidynamo theorem applies in the absense of the external magnetic field or a magnetic source. \cite{Seshasayanan14}   For convenience we plot spectral quantities in 3D XMHD in Fig.~\ref{3dxmhdplots} (see Ref.~\onlinecite{miloshevich17} and Table \ref{EvsIMHD} for more details).

We are led to conclude  that there may be barriers for finer-scale fluctuation amplifications (such as $k d_e \gg 1$). Also a natural conclusion could be that fluctuations of magnetic helicity $F$ are suppressed on the $d_e^{-1}$ scale.  Often times in this regime the electron MHD (EMHD) model is applied \cite{biskamp96,meyrand10,meyrandproc10} and so it is worthwhile comparing these models.  For the analysis,  what matters are integrals of motion, thus  we compare EMHD and inertial MHD ({which corresponds to Eqs.}~\eqref{2dxmhdrealequations} with $\rho_s = 0$) in   Table \ref{EvsIMHD}. It appears that the similarity is greater in 2D than in 3D. {Direct cascade of energy is also found in Ref.~\onlinecite{biskamp99}. However the model we use can also have non-zero electron temperatures ($\rho_s \ne 0$) that for some choice of parameters can lead to the inverse cascade of energy. }

\begin{table*}[htpb]
	\vspace{.25cm}
	\centering
	{\renewcommand{\arraystretch}{1.5}%
		{\setlength{\tabcolsep}{1.0em}
			\begin{tabular}{r|c|c|c}
				\hline
				Models & Energy & Magnetic-Helicity & Cross- \\
				\hline
				2D EMHD & $-\int d^2x\,\f{1}{2} (\psi^\star\nabla^2\psi + \phi^\star\omega)$ & $\int d^2x\,\f{1}{2}(\psi^\star)^2$ & $\int d^2x\,\phi^\star \psi^\star$ \\
				\hline
				2D IMHD & $-\int d^2x\,\f{1}{2} (\psi^\star\nabla^2\psi + \phi\,\omega)$ & $\int d^2x\,\f{1}{2}(\psi^\star)^2$ & $\int d^2x\,\omega\psi^\star$  \\
				\hline
				3D EMHD & $\int d^3x\,\f{1}{2} \bf{B}^\star\cdot \bf{B}$ & $\int d^3x\,\f{1}{2}\bf{A}^\star\cdot \bf{B}^\star$ &  \\
				\hline
				3D IMHD & $\int d^3x\,\f{1}{2} (V^2 + \bf{B}^\star\cdot \bf{B})$ & $\int d^3x\,\f{1}{2}(\bf{A}^\star\cdot \bf{B}^\star + d_e^2 \bf{V}\cdot \nabla\times \bf{V})$ & $\int d^3x\,\bf{V}\cdot \bf{B}^\star$ \\
				\hline
				3D XMHD & $\int d^3x\,\f{1}{2} (V^2 + \bf{B}^\star\cdot \bf{B})$ & $\int d^3x\,\f{1}{2}(\bf{A}^\star\cdot \bf{B}^\star + d_e^2 \bf{V}\cdot \nabla\times \bf{V})$ & $\int d^3x\,\lf({\bf{V}}\cdot {\bf{B}}^\star+\dfrac{d_i}{2}\bf{V}\cdot\nabla\times \bf{V}\ri)$ \\
				\hline
			\end{tabular}
		}
	}
	\caption{Comparison of the integrands for invariants in various extended MHD models. Notice that IMHD is normalized to the Alfven time-scale, while EMHD to a whistler time-scale $\tau_H = L^2\omega_{pe}^2/(c^2\Omega_e)$. In all cases the operator $*:= 1-d_e^2 \nabla^2$. }\label{EvsIMHD}
\end{table*}

2D IMHD and 2D EMHD can both be derived from 2D  XMHD in specific limits. The former, as already mentioned, is obtained after setting   to zero the out-of-plane components of the velocity and magnetic field. The latter is obtained by rescaling  
 the time with respect to the whistler time and  by retaining the leading order terms in the limit $d_i \gg 1$, where $d_i$ is the normalized ion skin depth. This comparison puts us in a position to discuss recent comments \cite{zhu17} regarding 2-fluid absolute 
 equilibrium states.\cite{miloshevich17,zhu14}  We agree with Ref.~\onlinecite{zhu17} that the qualitative picture of a direct cascade of the magnetic helicity is achieved in both 3D EMHD\cite{zhu14} and 3D IMHD;\cite{miloshevich17} however,  the details of spectral dependence are different, for instance in our model \cite{miloshevich17} we recover energy equipartition for  MHD.

When the effects of ion sound Larmor radius are included the eigenvalue analysis demonstrates that  NTS ($\alpha< 0$) are possible and we observe that in the low $k$ limit the total energy per wave-number $2\pi k \langle\calH (k)\rangle$ scales inversely with $k$ for the portion of inertial range, suggesting inverse cascade of energy (see Fig.~\ref{gyroonlyenergy}), as was first predicted by Onsager\cite{onsager49}  for  two-dimensional hydrodynamics. The inverse cascade of energy can also be inferred from the expression \eqref{hamexpectation} because $\beta$ is so large. Observed dependence of the invariants in this regime qualitatively agrees with the picture of the dual cascade obtained in drift wave two-field fluid models \cite{gang89,koniges91} and a gyrokinetic model \cite{zhu10} investigated later.  

Naturally, prior to   proceeding to the more general reduced extended MHD case like that of Ref.~\onlinecite{grasso16}, these predictions have to be confirmed by direct numerical simulations. For instance, there is evidence of broken ergodicity and coherent structures\cite{shebalin89,shebalin13} in MHD. Broken ergodicity is observed in many other physical systems including classical dipolar spin systems.\cite{miloshevich13} It is most suitable to consider a pseudo-spectral code  \cite{grasso99} to investigate whether the relaxation of the Fourier modes in MHD can occur. The advantages of using Galerkin methods in general involve accuracy and ``semiconservation`` of the integrals of motion.\cite{orszag71I}  Although for us it has an additional advantage since we are interested in the $k$-space behavior. Alternatively, since relaxation to equilibria subject to constraints is sought, it could be beneficial to apply recently developed symplectic/Poisson integration algorithms like the ones of Refs.~\onlinecite{kraus16,xiao2016}. This would also further justify the Hamiltonian treatment the problem has received. 

In closing we note that there  are many  plasma models where similar analysis can be performed. One of the candidates we intend to work with  in  the future is a special relativistic two-fluid model that was recently shown to possess Hamiltonian form.\cite{kawazura17}  This model can be applied in relativistic jets and laser fusion.
\acknowledgments
The work of GM and PJM was supported by the U.S. Department of Energy under Contract No.~DE-FG02-04ER-54742. We would like to thank Manasvi Lingam,   {Alexandros Alexakis} and   {Santiago Benavides}  for helpful discussions.
    \bibliographystyle{apsrev4-1}
    \bibliography{references}
\end{document}